\documentclass{article}
\usepackage[T1]{fontenc} 
\usepackage[utf8]{inputenc} 
\usepackage{ismir,amsmath,cite,url}
\usepackage{graphicx}
\usepackage{amsfonts}
\usepackage{booktabs}
\usepackage{color}
\def\lbd{}	
\usepackage{verbatim}
\usepackage{lineno}

\title{Mel-Band RoFormer for Music Source Separation}

\oneauthor
{Ju-Chiang Wang \hspace{0.5cm} Wei-Tsung Lu \hspace{0.5cm} Minz Won}
{Speech, Audio, and Music Intelligence (SAMI), ByteDance \\
{\tt\small \{ju-chiang.wang, weitsung.lu, minzwon\}@bytedance.com}}

\def\authorname{J.-C. Wang, W.-T. Lu, and M. Won}

\begin{document}

\maketitle
\begin{abstract}
Recently, multi-band spectrogram-based approaches such as Band-Split RNN (BSRNN) have demonstrated promising results for music source separation. In our recent work, we introduce the BS-RoFormer model which inherits the idea of band-split scheme in BSRNN at the front-end, and then uses the hierarchical Transformer with Rotary Position Embedding (RoPE) to model the inner-band and inter-band sequences for multi-band mask estimation. This model has achieved state-of-the-art performance, but the band-split scheme is defined empirically, without analytic supports from the literature. In this paper, we propose \emph{Mel-RoFormer}, which adopts the Mel-band scheme that maps the frequency bins into overlapped subbands according to the mel scale. In contract, the band-split mapping in BSRNN and BS-RoFormer is non-overlapping and designed based on heuristics. Using the MUSDB18HQ dataset for experiments, we demonstrate that Mel-RoFormer outperforms BS-RoFormer in the separation tasks of vocals, drums, and other stems. 


\end{abstract}
\section{Introduction}\label{sec:introduction}

Music source separation (MSS) \cite{rafii2018overview, mitsufuji2022music} aims to separate a music recording into musically distinct sources. Following the definition of the 2015 Signal Separation Evaluation Campaign (SiSEC) \cite{liutkus20172016}, the task is focused on the 4-stem setting: vocals, bass, drums, and other. 
The MUSDB18 dataset \cite{rafii2017musdb18} has been used to benchmark the performance. 

Different from CNN-based approaches \cite{chandna2017monoaural, kong2021decoupling, jansson2017singing} that make no assumptions on weighting different frequency bands, Band-Split RNN (BSRNN) \cite{luo2023music} directly splits the input frequency space into multiple subbands and models different subbands as a sequence. This multi-band approach has demonstrated promising results for MSS. In our recent work, we introduce the BS-RoFormer model \cite{lu2023music} which inherits the idea of band-splitting at the front-end. Then the model employs the hierarchical Transformer with Rotary Position Embedding (RoPE) to model the inner-band and inter-band representations as hierarchical sequences for multi-band mask estimation. Training a BS-RoFormer with MUSDB18HQ and 500 extra songs has achieved an average SDR of 11.99 dB, largely advancing the state-of-the-art performance of MUSDB18HQ. We submitted the system to the Music Separation track of Sound Demixing Challenge 2023 (SDX'23).\footnote{\url{https://www.aicrowd.com/challenges/sound-demixing-challenge-2023/}} Our system ranked the first place and outperformed the second best by a large margin in SDR \cite{fabbro2023sound}.
In ablation study \cite{lu2023music}, we demonstrate that RoPE is crucial in Transformer, and that a smaller BS-RoFormer model trained solely on MUSDB18HQ can also achieve very promising results, outperforming all existing systems that are trained without extra training data . 

From Psychoacoustics \cite{ballou2013handbook}, we learn that human auditory system tends to prefer higher resolution at lower frequencies, and is less sensitive at higher frequencies. This sets the basic principle when designing the band-split module in BS-RoFormer. However, such band-split scheme is defined empirically without analytic supports from the literature. In this paper, we explore the mel scale \cite{stevens1937scale}, which has a long history as the fundamental reference for acoustic feature design (e.g., MFCC and mel-spectrogram) in the field of audio signal processing. By replacing the band-split module with the so-called Mel-band projection module, we develop the \emph{Mel-RoFormer} model. In experiments, we show that Mel-RoFormer outperforms BS-RoFormer in the separation tasks of vocals, drums, and other stems. 

\begin{table*}[t]
\begin{tabular}{lcccccc}
 \toprule
 & Vocals & Bass & Drums & Other & Average & \# Param \\
 \midrule
 HDemucs \cite{defossez2021hybrid} & 8.04 & 8.67 & 8.58 & 5.59 & 7.72 \\
 BSRNN \cite{luo2023music} & 10.01 & 7.22 & 9.01 & 6.70 & 8.24 \\
 TFC-TDF-UNet-V3 \cite{kim2023sound} & 9.59 & 8.45 & 8.44 & 6.86 & 8.34 \\
 \midrule
 BS-RoFormer ($L$=6)  & 10.78 & 11.43 & 9.61 & 7.86 & 9.92 & 72.2M \\
 Mel-RoFormer ($L$=6) & 11.21 & 9.64  & \textbf{9.91} & 7.81 & 9.64 & 84.2M\\
 BS-RoFormer ($L$=9)  & 11.02 & \textbf{11.58} & 9.66 & 7.80 & \textbf{10.02} & 82.8M \\
 Mel-RoFormer ($L$=9) & \textbf{11.60} &  -  & 9.34 & \textbf{7.93} & - & 94.8M \\ 
 \bottomrule
 \end{tabular}
\centering
\caption{Comparison of different models (without extra training data) on MUSDB18HQ test set.}
\label{table:result}
\end{table*}

\section{Mel-Band Projection Module}
\label{sec:mel-band}

The Mel-band projection module relies on a mapping that projects relevant frequency bins to each specific band according to the mel scale, which is designed following a quasi-logarithmic function of acoustic frequency such that perceptually similar pitch intervals (e.g., octaves) have equal width over the full hearing range. 

\begin{figure}[t]
  \centering
 \centerline{\includegraphics[width=\columnwidth]{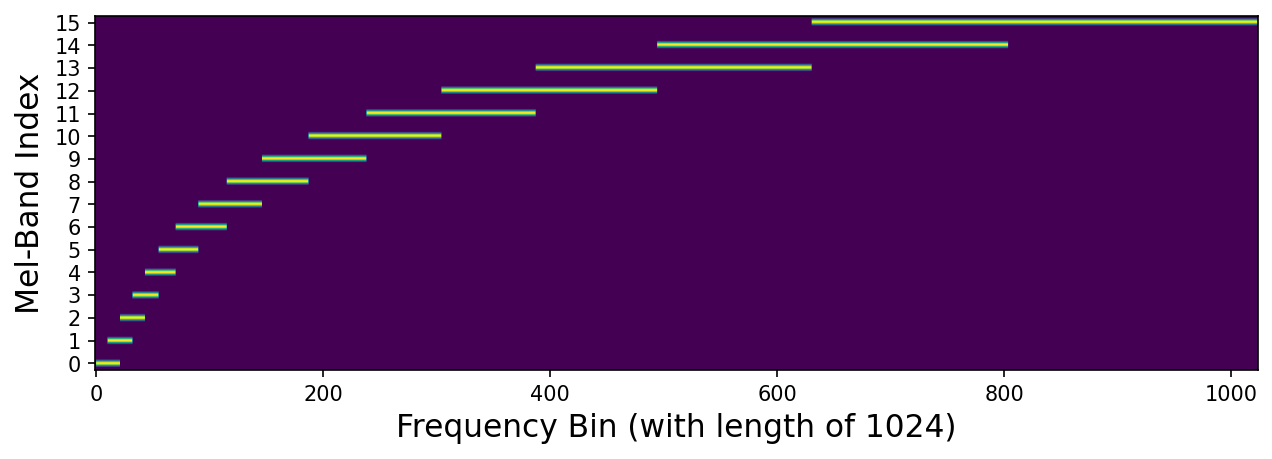}}
  \caption{The binary mapping between frequency bins and Mel-bands (with 16 bands). In this case, the frequency bins between 1 and 21 are projected into the 0-th Mel-band, the frequency bins between 11 and 32 are projected into the 1-th Mel-band, and so on. It can be seen that half of the frequency bins of the 0-th Mel-band overlaps the 1-th Mel-band, and that the bandwidth is larger at higher frequency. 
  }
  \label{fig:mel16}
\end{figure}

Given the number of Mel-bands for Mel-RoFormer, the center frequency of each Mel-band can be calculated on the mel scale. The width of a Mel-band is two times the distance between its center and its previous Mel-band's center. This makes the second half of a Mel-band overlaps its next Mel-band, and so forth until the last Mel-band. On the contrary, as a result of the band-split module in BS-RoFormer, the frequency ranges of different subbands are non-overlapping. Figure \ref{fig:mel16} illustrates an example of the Mel-band projection with 16 bands. In this case, a windows size of 2048 is used for FFT computation, so the length of frequency bins is 1024. During the multi-band mask estimation, each Mel-band representation is projected back to the original frequency space. Different from BS-RoFormer \cite{lu2023music}, the projected mask estimation values of the overlapped frequency bins are averaged accordingly to produce the final mask. Note that since we use the complex spectrogram as features, the Mel-band projection is applied to both real and imaginary values.

To retrieve the frequency-to-Mel-band index mapping, we utilize the implementation of Mel filter-bank in librosa \cite{mcfee2015librosa}, where  the mel-frequency replicates the behavior of the function in MATLAB Auditory Toolbox \cite{slaney1998auditory}. By calling \texttt{librosa.filters.mel} we obtain the mapping matrix with a triangle filter for each Mel-band. Then, we binarize this matrix by setting all non-zero values to 1 to discard the triangle filters. Such result yields the example in Figure \ref{fig:mel16}. Technically speaking, the Mel-band projection module can be seen as a learnable Mel filter-bank, since its MLP-layers serve as the mechanism to learn the filters.

\section{Experiment}
\label{sec:exp}

\subsection{Configuration}

Our experiment focuses on validating the effectiveness of Mel-RoFormer and if it can outperform the baseline BS-RoFormer. As training a larger model with more data takes a long time, we opt for smaller sizes of model configuration and use only MUSDB18HQ \cite{rafii2017musdb18} without adding any extra data. Specifically, we compare different models of using $L$=6 and $L$=9 for the RoPE Transformer block. We use 60 Mel-bands, as it is similar to the number of subbands, i.e., 62, adopted by BS-RoFormer. For deframing method, ``overlap \& average'' with a hop of half chunk is used for all models. All other configuration remains the same between Mel-RoFormer and BS-RoFormer \cite{lu2023music}. In terms of hardware, we use 16 Nvidia V100-32GB GPUs, and this leads to an effective batch size of 96 (i.e., 6 for each GPU) using accumulate\_grad\_batches=2.

\subsection{Results}

Table \ref{table:result} presents the results. We use the signal-to-distortion ratio (SDR) \cite{vincent2006performance} implemented by \texttt{museval} \cite{SiSEC18} as the evaluation metric. The median SDR across the median SDRs over all 1 second chunks of each test song is reported, following prior works. 
It is clear that the Mel-band projection can help the separation of vocals, improving the performance largely against the band-split module (e.g., by 0.43 dB and 0.58 dB for $L$=6 and $L$=9 models, respectively). This makes sense because the mel scale has been well proven to be useful in modeling human voices. Mel-RoFormer also outperform BS-RoFormer in the separation tasks of `drums' and `other' stems, but a deeper model ($L$=9) does not seem to help for drums. Qualitative analysis indicates that Mel-RoFormer can produce smoother vocal sounds with more consistent loudness. We will present more audio examples to attendees at the conference.

However, the Mel-band mapping is less successful for modeling the bass stem as compared to our band-split setting \cite{lu2023music}. We found the training progress became very slow when using Mel-RoFormer for bass, so we only report the result for $L$=6 model. Such observation is reasonable because bass is a unique instrument among the 4 stems that specifically focuses on low frequency. We also tried removing the overlapped frequency bins throughout the Mel-bands or using less Mel-bands, but the adjustments did not seem to help. This may indicate that the mel scale is an imperfect scheme to well characterize the timbres of bass.

\section{Conclusion}

We have shown that Mel-band projection is a promising scheme for multi-band MSS approaches for non-bass instruments. For future work, we plan to explore other supervised MIR tasks such as multi-instrument transcription \cite{lu2021spectnt, lu2023multitrack}, chord recognition, beat/downbeat tracking \cite{hung2022modeling}, and structure segmentation \cite{wang2022catch} using Mel-RoFormer.

\bibliography{mss}

\end{document}